\documentclass[runningheads]{llncs}
\usepackage[T1]{fontenc}
\usepackage{array}
\usepackage{graphicx}
\usepackage[inline]{enumitem}
\usepackage{wrapfig,lipsum,booktabs}
\usepackage{cite}
\usepackage{tikz}
\usetikzlibrary{shapes, quotes, arrows.meta, arrows, calc, shadows, positioning, fit, bending}
\usepackage{gastex}
\usepackage{wrapfig}
\usepackage{amsmath,amssymb,amsfonts}
\usepackage{algorithmic}
\usepackage{graphicx}
\usepackage{textcomp}
\usepackage{xcolor}
\usepackage{mathtools}
\usepackage{tikz}
\usepackage{centernot}
\usepackage{bussproofs}
\usepackage{verbatim}
\usepackage{latexsym}
\usepackage{pifont}
\usepackage{mathtools}
\usepackage{wrapfig}
\usepackage{hyperref}
\usepackage{booktabs}
\usepackage{makecell}
\usepackage{multirow}

\let\oldcenter\center
\let\oldendcenter\endcenter
\renewenvironment{center}{\setlength\topsep{6pt}\oldcenter}{\oldendcenter}
\allowdisplaybreaks
\begin{document}
\title{Probabilistic Model Checking of Temporal Interaction Dynamics in the Supreme Court}
\titlerunning{}
\author{Susmoy Das \and Arpit Sharma}
\authorrunning{S. Das and A. Sharma}
\institute{Department of Electrical Engineering and Computer Science\\Indian Institute of Science Education and Research Bhopal, India\\\email{\{susmoy18,arpit\}@iiserb.ac.in}\\}%
\maketitle

\begin{abstract}
The Supreme Court plays an extremely critical role in ensuring adherence to the rule of law and in strengthening the democracy. Due to this reason, modeling and analysis of small group interactions in the courtroom setting is an important task as it can help in understanding court decision-making. We apply probabilistic model checking for the modeling and analysis of temporal interaction dynamics in the context of the Supreme Court of the United States. We have used the transcripts of the oral arguments of cases from the Supreme Court for constructing a discrete-time Markov reward model (DTMRM). Next, we formulate interesting queries over interaction by using probabilistic computation tree logic (PCTL) and PCTL with rewards and verify them using a probabilistic symbolic model checker (PRISM). Our experimental results show that probabilistic model checking is very effective in identifying trends, hidden patterns, and how justices behave during the trials. These results not only provide valuable feedback to the justices but may also be used by the advocates and law students for finding better ways to present their arguments in the court.   

\keywords{Interaction analysis \and Markov chains \and Temporal patterns \and Logic \and Model checking \and Social sequence.}
\end{abstract}

\section{Introduction}\label{sec:intro}
Supreme Courts play a key role in modern democracies as they focus on cases of the greatest public and constitutional importance. They are responsible for upholding the rule of law and for ensuring that the rights of all citizens are preserved. It is, therefore, extremely important to study, model, and analyze the temporal dynamics of interactions \cite{cornwell} in the courtroom setting for understanding court decision making, and how justices and advocates behave during the trials. For example, it may be used to answer the following queries: how often do justices react to each other?, does the behavior of a justice change when he/she and the rest of the bench have opposite views on the case?, how would the bench react whenever an advocate expresses negative sentiment in his/her arguments?, and is there a necessary critical mass of self-conscious justices which can guide the resolution of a case?. Modeling temporal interactions and detecting behavioral patterns that emerge over time is a very complex and time-consuming task.

This paper presents a novel application of probabilistic model checking for modeling and analysis of temporal interaction dynamics in the context of the Supreme Court of the United States. Probabilistic model checking \cite{KB08} is an automated verification technique that is widely used for the evaluation of performance and dependability of information processing systems, e.g., randomized protocols. State-of-the-art probabilistic model checking tools such as Probabilistic Symbolic Model Checker (PRISM) \cite{PRISM1}, and Storm \cite{storm1,STORM} have been developed to model check Probabilistic Computation Tree Logic (PCTL) \cite{PCTL2,KB08} and PCTL with rewards \cite{Andova} on discrete-time Markov chains (DTMCs) \cite{KB08}. We have used the transcripts of the oral arguments of cases from the `Supreme Court Oral Arguments Corpus' \cite{dataset1,dataset} for constructing a discrete-time Markov reward model capturing the interaction behavior in the courtroom scenario. A discrete-time Markov model has been used because the group interactions involving spontaneous and unscripted speech are known to follow the Markovian property \cite{murrayold,oana1}. Every case has been converted into a trace, i.e., a sequence of utterances where each utterance is represented as a state, i.e., $10$-tuple of this model. In other words, states have been used to store important information about every utterance of the case, e.g., speaker name, speaker side, votes side, utterance type, and utterance sentiment. The state transition probabilities have been computed directly from the transition counts in the corpus. Next, we specify interesting properties over behavior using PCTL and PCTL with rewards and verify them on the PRISM model checker.

Our experimental results show that probabilistic model checking of courtroom behavior can be very effective in analyzing the interactions and identifying trends and hidden patterns. For example, our results demonstrate that except for one justice, all the other justices tend to participate in the trial right from the start. Similarly, it was observed that cases where a unanimous verdict was passed, were not only shorter in length than average, but the justices also had limited involvement (interventions) in these cases. In contrast, cases where the bench is divided, i.e., $5-4$ are not only longer than average but the expected number of interventions from the justices is also higher than average. Another interesting observation is that like-minded jurors tend to jointly intervene the advocates of the opposite side during the trials. Our results also show that rebuttals usually go unintervened, but for certain cases, they may result in long one-on-one conversations between a justice and an advocate. Finally, our results also empirically validate the claims made by popular newspaper articles, e.g., \cite{silent_tom2,silent_tom3}. This type of analysis can be very helpful for the justices as it provides them valuable feedback to enable better coordination, information about specific issues that need to be handled, and to find out how they actually vote. Additionally, it may also be useful for the advocates and the law students as they can identify better ways to present their arguments during the trials. Courtrooms across the globe may also model their cases, and compare/correlate their results with our findings.

To the best of our knowledge, this is the first work on modeling and analysis of temporal interaction dynamics in the courtroom setting using probabilistic model checking. We believe this work would encourage researchers from different disciplines, e.g., computer science, computational social science, psychology, and linguistics to jointly focus on the 
problems of small group interactions \cite{cornwell} in the judiciary and other related domains.     

\paragraph{Organization of the paper.} The rest of the paper is organized as follows. Section~\ref{sec:related} briefly discusses the related work. Section~\ref{sec:prelim} recalls the basic concepts of DTMCs, probabilistic logics, and also discusses the ConvoKit dataset. Section~\ref{sec:model} discusses the procedure for building a Markovian model followed by Section~\ref{sec:method} which presents the reward structures and filters. Section~\ref{sec:queries} defines the queries and analyses them. Finally, Section~\ref{sec:conc} concludes the paper and provides pointers for future research.

\section{Related Work}\label{sec:related}
In \cite{relatedcourt1}, authors investigate the differences in discourse styles, e.g., information density for various stakeholders involved in the Swedish court hearings. The results have been interpreted using speech accommodation theory. In \cite{relatedcourt4}, authors use the transcripts from the U.S. Supreme Court to explore the hypothesis that discourse markers and personal references are important features in models exhibiting the turn-taking behavior of participants. In \cite{relatedcourt3}, authors study and measure the pair-wise similarity between adjacent filled pauses in justice-lawyer pairs. Their results show that the mean difference between these pauses was smaller for those justice-lawyer pairs where the justice gave a favorable vote as compared to the pairs where the justice gave a non-favorable vote. In \cite{dataset}, authors show how variations in linguistic styles can provide information about power differences within social groups. For this study, authors have used discussions between editors on Wikipedia and oral arguments of the U.S. Supreme Court. In \cite{relatedcourt5}, authors demonstrate using a predictive model that justices implicitly reveal their leanings during oral arguments, even before arguments and deliberations have concluded by extracting emotional arousal using vocal pitch from audio transcripts of the proceedings of the Supreme Court of the U.S.

Unlike these research papers, we focus on end-to-end modeling of the behavior, which enables us to reason about temporal dynamics in peer interaction events in the courtroom setting. The idea of using Markov reward models for studying group dynamics was originally proposed by Murray in \cite{murrayold}. In this paper, a value iteration algorithm has been proposed to estimate the value of every state. Andrei et al. \cite{oana1} have extended this line of research by empirically demonstrating the expressiveness of probabilistic temporal logic and model checking for the analysis of group dynamics in social meetings. 

\section{Preliminaries}\label{sec:prelim}
This section recalls the basic concepts of discrete-time Markov chains (DTMCs) with finite state space and probabilistic computation tree logic (PCTL) used for specifying properties over these models.
\paragraph{\textbf{DTMC\cite{dtmc1,KB08}}:}A DTMC is a tuple $\mathcal{D}=(S,AP,\mathcal{P},s_{0},L)$ where: (a) $S$ is a countable, nonempty set of states; (b) $AP$ is the set of atomic propositions; (c) $\mathcal{P}$ is the transition probability function satisfying $\mathcal{P}: S \times S \rightarrow [0,1]$ s.t.\ $\forall s \in S$: $\sum_{s' \in S}\mathcal{P}(s,s')=1$; (d) $s_{0}$ is the initial state; and (e) $L: S \rightarrow 2^{AP}$ \normalfont{is a labeling function.}
$\mathcal{D}$ is called finite if $S$ and $AP$ are finite. Let $\rightarrow=\{(s,p,s')\mid \mathcal{P}(s,s')=p$\textgreater$0\}$ denote the set of all the transitions for a DTMC $\mathcal{D}$. We denote $s\xrightarrow{p}s'$ if $(s,p,s')\in\rightarrow$. A sequence of states $s_0,s_1,s_2,\ldots$ where $\mathcal{P}(s_i,s_{i+1})>0$ $\forall i$ is an infinite path in a DTMC. We denote a path by $\pi$. A finite path is of the form : $s_0,s_1,s_2,\ldots,s_n$ where $\mathcal{P}(s_i,s_{i+1})$\textgreater$0$ for $0\le i< n$. The length of a finite path, denoted by $len(\pi)$ is given by the number of transitions along that path. The length of the finite path given above is $len(\pi)=n$. For an infinite path $\pi$, we have, $len(\pi)=\infty$. We denote the $n$-th state along a path $\pi$ by $\pi[n-1]$ ($\pi[0]$ denotes the first state from which the path starts). Let $Paths(s)$ denote the set of all infinite paths starting in $s$. Let $Paths_{fin}(s)$ denote the set of all finite paths starting in $s$. Let $s_{0},\ldots,s_{k}\in S$ with $\mathcal{P}(s_{i},s_{i+1})>0$ for $0\le i<k$. $Cyl(s_{0},\ldots,s_{k})$ denotes the \emph{cylinder set} \cite{KB08,Sharma1} consisting of all paths $\pi\in Paths(s_{0})$ s.t.\ $\pi[i]=s_{i}$ for $0\le i\le k$. Let $\mathcal{F}(Paths(s_{0}))$ be the smallest $\sigma$-algebra on $Paths(s_{0})$ which contains all sets $Cyl(s_{0},\ldots,s_{k})$ s.t.\ $s_{0},\ldots,s_{k}$ is a state sequence with $\mathcal{P}(s_{i},s_{i{+}1}) > 0$, ($0 \le i < k$). The probability measure $\Pr$ on $\mathcal{F}(Path(s_{0}))$ is the unique measure defined by induction on $k$ in the following way. Let $\Pr(Cyl(s_{0}))=1$ and for $k > 0$ : $\Pr(Cyl(s_{0},\ldots,s_{k},s')) = 
		\Pr(Cyl(s_{0},\ldots,s_{k})) \cdot \mathcal{P}(s_k,s')$.
\paragraph{\textbf{PCTL\cite{PCTL2,KB08}}:}PCTL is a probabilistic branching-time temporal logic that allows one to express the probability measures of satisfaction for a temporal property by a state in a DTMC. The syntax is given by the following grammar where $\Phi,\Phi',\ldots$ range over PCTL state formulae and $\Psi,\Psi',\ldots$ range over path formulae:\\
\begin{itemize*}
	\item State Formulae:~~$ \Phi ::== \textbf{true} \mid a \mid $ $\lnot$ $\Phi \mid \Phi \wedge \Phi' \mid \mathcal{P}_{J}(\Psi) $, for some $a\in$AP\\
	\item Path Formulae :~~$ \Psi ::== \textbf{X} \Phi \mid \Phi\textbf{U}\Phi'$
\end{itemize*}\\
where $J\subseteq[0,1]\subset\mathbb{R}$ is an interval. Satisfaction of a PCTL state formula $\Phi$ by a state $s$ or a path formula $\Psi$ by a path $\pi$, notation, $s\models\Phi$ or $\pi\models\Psi$ is defined inductively by :\\
\begin{itemize*}
	\item$s\models\textbf{true}\text{ always}$;
	\item$s\models a\text{ iff }a\in L(s)$;
	\item$s\models\lnot\Phi\text{ iff }s\not\models\Phi$;\\
	\item$s\models\Phi \wedge \Phi'\text{ iff }s\models\Phi\text{ and }s\models\Phi'$;\\
	\item$s\models \mathcal{P}_{J}(\Psi)\text{ iff }\sum_{\pi\in Paths(s),\pi\models\Psi}Pr(\pi)\in J$;
	\item$\pi\models\textbf{X}\Phi\text{ iff }\pi[1]\models\Phi$;\text{ and}\\
	\item$\pi\models\Phi\textbf{U}\Phi'\text{ iff }\exists k\ge 0, \text{ s.t.\ }\pi[k]\models\Phi'\text{ and }\forall 0\le i< k,\pi[i]\models\Phi$.
\end{itemize*}
\subsection{The Supreme Court Oral Arguments Corpus}\label{sec:corpus}
The `Cornell Conversational Analysis Toolkit (ConvoKit)' \cite{dataset1,dataset} contains tools that are capable of analyzing conversations and the social interactions embedded within. It provides intuitive and user-friendly abstractions that enable one to both represent and manipulate conversational data. Several interesting conversational datasets are part of this toolkit, and the `Supreme Court Oral Arguments Corpus' \cite{website-ckit} that we have used in our work is also part of this collection. The dataset contains a collection of cases from the Supreme Court of the United States, along with the transcripts of the oral arguments of these cases.
\paragraph{Dataset Details:}
This dataset is split into different years spanning $1955$ to $2019$. Each case can be identified uniquely and has the following attributes:
\begin{itemize}
	\item Speaker-level Information: contains a unique \verb|ID| for each speaker, \verb|name| and \verb|type|, which could either be a justice, an advocate, or an unknown.
	\item Conversation-level Information: contains a unique \verb|ID|, the \verb|case_ID| giving information about the case to which it belongs, an \verb|advocates| dictionary containing the details of each advocate involved in the case and their \verb|role|, i.e., the side which they represented and if they were \textit{`amicus curiae'} or not, \verb|votes_side| which is again a dictionary where each entry lists the side for which a particular justice voted in that session, and \verb|win_side| which contains the information of the outcome of the case.
	\item Utterance-level Information: it contains a unique \verb|ID|, \verb|speaker|-information, \verb|conversation_ID|, \verb|case_ID|, \verb|speaker_type| which gives the information if the speaker is a justice or an advocate, \verb|side| the side of the speaker, the utterance \verb|text| and other utterance related information.
\end{itemize}
We have considered $125$ cases from this dataset collected over the span of two years, i.e., $2018$ and $2019$ as the bench of the justices was identical in both these years. For these subsets of cases, we also had the voting information available for more than half of the bench.

\section{Model Construction}\label{sec:model}
We pre-process the dataset to extract and keep the information that is required for modeling the behavior and discard all the other attributes which are not used in our analysis. We construct an interim `CSV' file. For every case, we convert each utterance to a row in the CSV file which would preserve information about its parent case. Next, we consider every case as a trace, i.e., a sequence of utterances where each utterance is a row of the CSV file. In other words, we have a collection of $125$ traces that were used for constructing the DTMC model. Each row has been converted into a tuple which is as follows:
\begin{enumerate}
\item Speakers' name: the name of the justice or `nreq' in case of the advocates;
\item Speakers' side: `JJ'- for a justice, `PP'- for the petitioner, `RR'- for the respondents, and `SU'- side unknown;
\item AC-value: `ACYES' if the speaker is an amicus-curiae, `ACNO' otherwise;
\item Win-side: takes the value $0$ if the case was decided in favor of the respondents, $1$ if it was in favor of the petitioners;
\item Votes-side: is a $9$-tuple representing the individual votes of the nine justices, where $0$ at $i^{th}$ position denote that the $i^{th}$ justice voted for the respondents, $1$ denotes vote in favor of the petitioner and "IU" means that the information is unknown. Here, the sequence used for the justices is as follows: $J_1$, $J_2$, $J_3$, $J_4$, $J_5$, $J_6$, $J_7$, $J_8$ and $J_9$.
\item Utterance-type: (a) Opening: The chief justice opens the case. (b) Closing: The chief justice closes the case. (c) Conclopening: The chief justice thanks one speaker and gives a turn to the next speaker. (d) Intervening: A justice intervenes a speaker. (e) Normal: Any utterance by an advocate not during the rebuttal. (f) Rebuttal: An utterance by an advocate during the rebuttal.
\item End of Utterance-type: (a) MS- Utterance finished in the middle of a sentence. (b) SF- Utterance finished with a sentence being completed. (c) LG- Utterance finished with laughter (d) NR- Utterance finished with no response.
\item Utterance-sentiment: Sentiment associated with the utterance, i.e., positive, negative, or neutral (no-sentiment).
\item Utterance-length: the length of the utterance, long (LoU) or short (SoU).
\item Number of pauses: the number of pauses in an utterance, i.e., more (MP) or less (LP).
\end{enumerate}

Here, `Intervening' can take one of the following tags: (a) Intervening(PP/RR) refers to the side that was being intervened (b) InterveningPPRE denotes that the petitioner is intervened during the rebuttal, (c) InterveningAC(PP/RR) is similar to (a) with the additional property that the speaker is an amicus curiae and (d) Intervening(PP/RR)JJ represents that two or more justices are intervening one after another. For example, a state tagged 'InterveningPPJJ', would require that `Intervening', 'InterveningPP', and `InterveningPPJJ' are all true in that state.

These components jointly constitute the state space of our Markovian model. Furthermore, we introduce two new states: `Initial' is the initial state of the model, which with equal probability, goes to each unique `opening' state of a case. Similarly, all the `closing' states reach a single `Final' state with probability $1$. The state transition probabilities have been calculated using the transition counts obtained from the dataset. For the utterance sentiment, we have used the Natural Language Toolkit (NLTK) sentiment analysis library \cite{nltk} which calculates the compounded score for each statement in the utterance. Finally, we take the average over all the sentences of an utterance and determine the value of utterance-sentiment. In the case of the last two tuples, for each speaker type, we calculated the average lengths of their utterances and also the average number of pauses used in these utterances. This allows us to identify a threshold that can be used to classify an utterance as either long or short. Similarly, we can also classify an utterance as either with more pauses or with less pauses.    

A typical courtroom proceeding begins with the chief justice opening the session and presenting the floor to the petitioner to present their argument. The argument is intervened by the bench of justices following which the chief justice invites the respondents side to present their argument which is also intervened by the justices. In the end, the petitioner gets a chance to present a rebuttal argument following which the chief justice closes the session and submits the case. Note that each side may have multiple arguments during a trial, but the broader framework remains the same. Additionally, there may be arguments by the amicus curiae (`friend of the court'), who is/are permitted to assist the court by offering additional information, perspective, and expertise about the case. An amicus curiae may or may not support any side.
\begin{example}
Consider the following state: $(J_6,JJ,ACNO,0,0,1,0,0,1,0,0,1,0),\\InterveningPP,MS,Neg,LoU,MP)$. This is a state where justice $J_6$ intervened the petitioner in a case where the respondents won. The utterance was long, and the sentiment was negative. The utterance ended mid-sentence and had more pauses. Here, it can be observed that justices $J_2$, $J_5$, and $J_8$ have voted for the petitioner, and the rest of the justices voted for the respondents. 
\end{example}
\begin{example}
We show a snapshot of our Markovian model in Fig.~\ref{fig:e1}. Here, the state $(NREQ,PP,ACNO,1,(1,1,1,1,1,1,1,1,1),Normal,LG,Pos,SoU,LP)$ is an utterance by a member from the petitioner side who is not an amicus curiae and the argument is not a rebuttal argument. The utterance was small with less number of pauses that ended with laughter. This is a case where the petitioner won unanimously. This utterance was immediately followed (amongst others) by an intervention from $J_6$ with probability $\frac{1}{7}$, and an intervention from $J_2$ with probability $\frac{3}{7}$.
\begin{figure}[h]
    \centering
    \includegraphics[width=\textwidth]{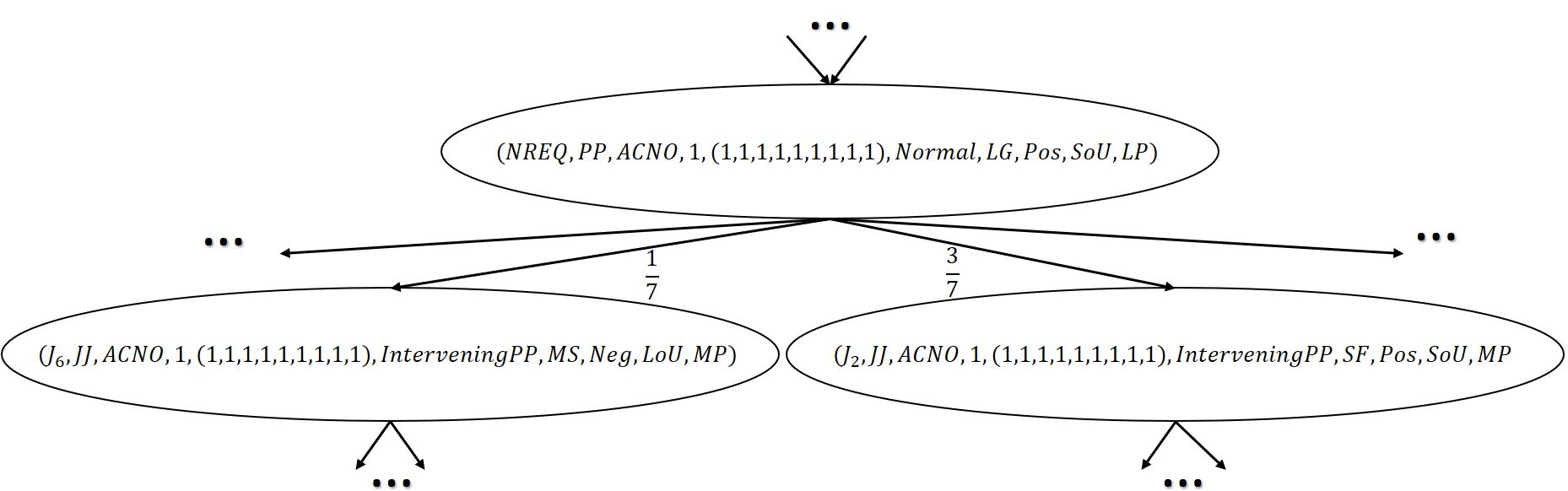}
    \caption{Snapshot of the Model}
    \label{fig:e1}
\end{figure}
The intervention by justice $J_6$ was a long utterance that ended mid-sentence, had a negative sentiment and more pauses. In contrast, the intervention by $J_2$ was a short utterance with a positive sentiment. The rest of the model can be interpreted in a similar fashion.
\end{example}

\section{Reward Structures}\label{sec:method}
For the modeling and verification, we use the PRISM model checker \cite{PRISM1}, which is widely used for the design and analysis of systems that exhibit probabilistic behavior. PRISM allows us to capture the system behavior using its modeling language and specify interesting properties using PCTL logic. Additionally, PRISM also allows the operator $\mathcal{P}=?$, which computes the probability of satisfying a property.

PRISM also allows us to augment the DTMC models with rewards, which are non-negative real-valued quantities assigned to states \cite{Andova,rewards2}. The resulting models are known as discrete-time Markov reward models (DTMRMs). We can define multiple reward structures on the same model \cite{Andova,MRMC}. This enables us to specify interesting reward-based properties using PCTL with rewards, which is also supported by PRISM. Reward-based properties can be broadly classified into two groups: 1) Instantaneous properties - the expected value of the reward at some time point, 2) Cumulative properties - the expected cumulated reward over some period. For example, $R\{r\}=?[C^{\le n}]$ computes the expected reward cumulated up to time-step $n$. Similarly, $R\{r\}=?[F~\Phi]$ computes the expected reward cumulated before reaching the set of states which satisfy $\Phi$.

PRISM uses filters to verify properties that are valid when initiated from a collection of states which meet specific conditions. In this paper, we use two distinct types of filter operators: (a) The filter $(state,\Phi,clause)$ assesses the satisfaction of a state-based formula ($\Phi$) from a state uniquely identified by the Boolean-valued expression ($clause$). (b) The filter $(avg,\Phi,clause)$ calculates the average across all the states where the condition $clause$ holds true. We can replace $avg$ by $min$ or $max$ to get the corresponding minimum and maximum values, respectively from the states where the condition $clause$ holds true.  

For the sake of convenience, we introduce the following short forms:
\begin{itemize}
	\item \verb|jname| can be replaced with the names of any of the justices,
    \item \verb|side| can be replaced with any of the speakers side,
    \item \verb|type| can be replaced with any of the end of utterance-type,
	\item \verb|sentiment| can be replaced with any of the sentiments, and
	\item \verb|pauses| can be replaced with MP or LP.
\end{itemize}
Next, we define the following reward structures on the model:
\begin{enumerate}
	\item \verb|n_step| assigns a value of $1$ to each state and is used to compute the number of steps (time);
	\item \verb|r_jname_intervening| assigns a value of $1$ to each state where \verb|jname| and `Intervening' is true. It is used to calculate the number of interventions by a particular justice;
	\item \verb|r_JJ_intervening_side| assigns a value of $1$ to each state where `JJ' and `Intervening' is true for a particular \verb|side|. It is used to calculate the total number of interventions by all the justices towards a particular side;
	\item \verb|r_jname_intervening_side| assigns a value of $1$ to each state where \verb|jname| and `Intervening' is true for a particular \verb|side|. It is used to calculate the number of interventions by a particular justice \verb|jname| towards a \verb|side|; and
	\item \verb|r_type| assigns a value of $1$ to each state where \verb|type| is true. It calculates the number of occurrences of each utterance type during a trial.
\end{enumerate}
One can suitably develop more reward structures based on the properties that need to be analyzed and verified.

\section{Experimental Results and Analysis}\label{sec:queries}
In order to analyze the interaction, we formulate several interesting properties/queries using PCTL and PCTL with rewards, and model check these queries on the PRISM toolset. The PRISM model and all the PRISM queries (PCTL properties and PCTL with reward properties) can be found on the Zenodo repository\footnote{\url{https://doi.org/10.5281/zenodo.8166141}}. Our PRISM model has $10099$ states and $24534$ transitions including a unique initial state and a unique final state. Throughout this section, we use $j$ in the queries to denote an integer variable which can take values ranging from $0-10098$ (number of states in our model). 

\subsubsection*{1. Queries for validating the model.}    
    \textbf{(a)} What is the expected number of steps required for the first involvement of each role during a proceeding?
		\begin{center}
        \scriptsize
			\verb|R{"n_step"}=?[F "side"]|
		\end{center}
The result of this query for the justices would be $1$ (as the chief justice opens the hearing), for the petitioner it is $\sim 3$ as the chief justice offers the floor to the petitioner. For respondents, it is $105$ as they participate at a later stage of the trial. The formula \verb|R{"n_step"}=?[F "CLOSING"]| will provide us with the expected duration of a proceeding, and the result is $\sim 234$. This shows that the respondents are offered the floor roughly halfway through a trial, as some time also needs to be allocated for the rebuttal towards the end of the trial. This is also expected in a courtroom where both sides should get the same amount of time for presenting their arguments.    
	
    \textbf{(b)} What is the average probability of an event where utterances from the advocates of opposite sides immediately follow each other?
		\begin{center}
        \scriptsize
			\verb|filter(avg,P=?[X ("side")],"side")|
		\end{center}
As advocates from one side never directly interact with the advocates of the other side during the trial, it is expected that the value of this query should be $0$. 
This is indeed the case for the following query: \verb|filter(avg,P=?[|\verb|X ("PP")],"RR")| and its counterpart. 

    \textbf{(c)} What is the average probability of an event where the model has erroneously skipped the justice name or not abstracted the name of the advocate?
		\begin{center}
        \scriptsize
			\verb|filter(avg,P=?[F ("NREQ"&"JJ")],"INITIAL")|=$0.0$~\\
			\verb!filter(avg,P=?[F ("NREQ"&("PP"|"RR"))],"INITIAL")!=$1.0$
		\end{center}
These queries ensure that there are no states in the model where the name of the justice is missing or the name of an advocate has been stored.
	
    \textbf{(d)} What is the probability that the current state and the next state of the model are the same?
		\begin{center}
    \scriptsize
			\verb|filter(max,P=? [X(x=j)],(x=j))|
		\end{center}
The result of this query confirmed our expectations that no two utterances can have the same set of values for all the attributes. This is due to the fact that utterances are segregated at those instances where the value of some attribute changes. The filter returned no states with non-zero probabilities, which implies that the probability is $0$.
 
    \textbf{(e)} What is the probability that a trial ends where only one side has presented its argument?
	The property can be encoded in PRISM as follows:
	\begin{center}
    \scriptsize
		\verb|P=?(!("PP")U("CLOSING"))| and \verb|P=?(!("RR")U("CLOSING"))|
	\end{center}
As expected, the value for these queries is $0$ as every proceeding must have at least one argument from both the sides. 
 
	\textbf{(f)} What proportion of cases went in the favor of the petitioner/respondents?
	\begin{center}
    \scriptsize
		\verb|P=?[(X("FP"))U("FP")]|=0.58~~\verb|P=?[(X("NFP"))U("NFP")]|=0.42
	\end{center}
We augment these queries with an additional clause for validating the model. The set of queries and their values are given below:
	\begin{center}
    \scriptsize
		\verb|filter(avg,P=?[X ("FP")],"NFP")|=0~~\verb|filter(avg,P=?[X ("FP")],"NFP")|=0
	\end{center}
From the first set of queries, it can be observed that the petitioner has a slight edge in winning the cases as compared to the respondents. The second set of queries assures us that the model does not mix states from cases having different judgments. 
 
    \textbf{(g)} What proportion of the judgments went in favor of the petitioner/respondent for each justice? This query can be encoded in PRISM as follows:
	\begin{center}
    \scriptsize
		\verb|P=?[(X"jname0")U("jname0")]|~~\verb|P=?[(X"jname1")U("jname1")]|
	\end{center} 
As mentioned earlier, for a small number of cases, the votes for some of the justices were unknown, and therefore a small probability mass is seen missing while classifying their judgments, viz., $J_6$ and $J_7$ ($1\%$), and $J_9$ ($7.4\%$). As expected, the results of this query also followed the same trend where all the justices have a slight inclination towards giving the decision in favor of the petitioner (except $J_6$). We also calculated the following: what proportion of judgments were given with a unanimous verdict? The result of this query is $31\%$ where $18\%$ were in favor of the petitioner and the rest was given in favor of the respondents. In order to further check the correctness of our model, we calculated if a state could be reached where a justice favors the respondent from a state where the same justice favors the petitioner. The corresponding query is as follows: \verb|filter(avg,P=?[(X"jname|\verb|0")],"jname1")| and \verb|filter(avg,P=?[(X"jname1")|\verb|],"jname0")|. As expected, the result is $0$ for each justice.
    
	\textbf{(h)} What are the expected occurrences of each utterance type during the course of a proceeding? This query can be formalized in PRISM as follows:
		\begin{center}
        \scriptsize
				\verb|filter(avg,R{"r_type"}=?[F "FINAL"],"INITIAL")|
			\end{center}	   
         As expected, the model has correctly captured the typical behavior of a courtroom proceeding where there will be a single instance of `Opening' and `Closing' (query result $=1$). `Conclopening' refers to the utterance where the chief justice thanks the advocate and moves to the next speaker. As both the sides participate in all the trials and this is followed by at least one rebuttal, the chief justice will need to thank the speakers at least twice. Hence, there has to be a minimum of 2 instances of `Conclopening' (query result $=3$), and at least one instance of `Rebuttal' (query result $=4$). The other occurrences are either `Normal' utterances (query result $=106$) or `Intervening' (query result $=116$) type utterances. The sum of each utterance type is $231$.
    
\subsubsection*{2. Sentiment Analysis of the environment in the courtroom.}\mbox{} \\
The first set of queries refers to the long-run probabilities of being in a state of `Positive/Negative/No-Sentiment' which can be encoded as follows:
     \begin{center}
    \scriptsize
		\verb|S=?["Pos"]|~~\verb|S=?["Neg"]|~~\verb|S=?["Neu"]|
     \end{center}
These queries result in the following values: $0.44$, $0.19$ and $0.37$, respectively. This indicates that the courtroom sentiment is either mostly positive or neutral, with occasional negativity. We can modify these queries to include the roles as well which will provide us with the long-run probabilities for each role with a given sentiment. The modified queries are: \verb |S=?["role"&"sentiment"]|. From these queries, we see that justices are more likely to be neutral ($0.2$) or positive ($0.2$) than negative ($0.1$). The petitioners/respondents mostly express positive sentiments ($0.1$), followed by no sentiments ($0.075$) and very few instances of negative sentiments ($0.025$). We can conclude that the overall sentiment of the courtroom is positive, and justices tend to express neutral or positive sentiments most of the time. In order to check if the stakeholders involved in a trial try to maintain a positive outlook inside the courtroom, we look at the following set of queries and their corresponding values:
	\begin{center}
    \scriptsize
		\verb|filter(avg,P=?[X("Pos")],"Pos")|=$0.51$~~\verb|filter(avg,P=?[X("Pos")],"Neg")|=$0.44$\\
		\verb|filter(avg,P=?[X("Neg")],"Neg")|=$0.23$~~\verb|filter(avg,P=?[X("Neg")],"Pos")|=$0.15$
	\end{center}
These values suggest that the stakeholders try to not get stuck in the negativity, and make efforts to get out of it. For example, if we are in a state with positive sentiment, then the chances of moving towards positivity ($0.51$) is much higher than negativity ($0.15$). Similarly, if we are in a state with negative sentiment, then the chances of coming out from it ($0.44$) are higher than continuing with the negativity ($0.23$).  	

\subsubsection*{3. What is the expected number of steps before each justice participates during a trial?}
The corresponding queries can be formalized as follows:
	\begin{center}
    \scriptsize
		\verb|R{"n_step"}=?[F ("jname")]|
	\end{center}
Note that, we need to modify the query for the chief justice because otherwise, the value is always going to be $1$ as he opens the session. The modified query is as follows: \verb|R{"n_step"}|\verb|=?[F ("jname"&!("OPENING"))]|. The average value for all the justices other than $J_2$ is $33$. The maximum number of expected steps is $50$ for $J_9$. This shows that all the justices get themselves involved in the case right from the start and, latest by the $50^{th}$ utterance which is $20\%$ of a trial. Justice $J_2$ is an exception, where the expected number of steps for the first involvement is $1134$.

\subsubsection*{4. What is the expected number of interventions by each justice during a trial?} The above query can be translated into PRISM as follows:
	\begin{center}
        \scriptsize
		\verb|filter(avg,R{"jname_INTERVENING"}=? [F "FINAL"],"INITIAL")|
		\verb|filter(avg,R{"jname_INTERVENINGside"}=?[F "FINAL"],"INITIAL")|
	\end{center}
We have observed that $J_6$ intervenes the most whereas $J_3$ and $J_2$ intervene much lesser than the average. Interestingly, for $J_2$, the value is $\sim0.8$ which means that there are proceedings where $J_2$ has remained silent throughout the trial. Additionally, we have also observed that the collective share of interventions towards the petitioners and the respondents side is very even ($6/7$). This indicates that the judiciary is free and fair. 

\subsubsection*{5. Analyzing the behavior of a justice in cases where his/her vote was the only one in favor of the losing side.}
This situation took place for six of the nine justices from the bench. The expected number of interventions by a justice can be seen from Table~\ref{tquery9}.

	\begin{table}[h]
		\scriptsize
		\caption{Interventions during normal trials to those where the judgment was $8-1$.}
		\label{tquery9}
		\centering
		\begin{tabular}{ | c | c | c | c | c | c | c | c | c | c |}
			\hline
			\multirowcell{2}{Justices} & \multicolumn{3}{|c|}{All cases} & \multicolumn{3}{|c|}{$8-1$ Cases} & \multicolumn{1}{|c|}{Justice} & \multicolumn{1}{|c|}{Outcome} & \multicolumn{1}{|c|}{\% Change}\\
			\cline{2-7}   &   Total  &  PP  &  RR  &  Total  &  PP  &  RR  &  vote  &   &  In favor \\
			\hline
			$J_6$ & 21 & 13 & 8 & 20 & 17 & 3 & 0 & 1 & -62.5\%\\
			\hline			
			$J_5$ & 13 & 6 & 7 & 13 & 12 & 1 & 0 & 1 & -85.7\%\\
			\hline			
			$J_8$ & 14 & 5 & 9 & 12 & 11 & 1 & 0 & 1 & -88.9\%\\
			\hline			
			$J_3$ & 7 & 4 & 3 & 8 & 5 & 3 & 1 & 0 & 25.0\%\\
			\hline			
    		$J_4$ & 18 & 8 & 10 & 4 & 2 & 2 & 1 & 0 & -75.0\%\\
			\hline			
			$J_2$ & 0.8 & 0.4 & 0.4 & 0 & 0 & 0 & 1 & 0 & -100\%\\
			\hline			
			$J_2$ & 0.8 & 0.4 & 0.4 & 0 & 0 & 0 & 0 & 1 & -100\%\\
			\hline					
		\end{tabular}
	\end{table}
 
 We see that justices tend to intervene less (on average it decreases by $\sim 65\%$). For example, in all the cases where only $J_6$ voted for the respondents, the expected number of interventions for the RR was only $3$ as compared to the expected number of interventions by $J_6$ for all the cases which is $8$. Similarly, $J_8$ voted for the respondents, and therefore expected number of interventions for the RR was $1$ as compared to the expected number of interventions over all possible cases which is $9$. Note that $J_3$ is an exception, where the expected number of interventions is increased by $1$ when the justice has voted for the petitioner. For the top three justices in the same table, we also see a rise in the expected number of interventions for the PP (the other side) when the justices voted for the RR.

\subsubsection*{6. Analyzing the changes in the behavior of the justice(s) for special cases.}
\textbf{(a)} Cases with unanimous judgment: In these experiments, we try to study the behavior of the justice(s) for cases where the bench had given a unanimous decision. We have compared the expected number of interventions for the cases where the decision is unanimous with the average over all the cases. An example PRISM query is shown below:
	\begin{center}
		\scriptsize
		\verb|filter(avg,R{"r_INTERVENING"}=?[F "FINAL"],"OPENING"&"(0,0,0,0,0,0,0,0,0)")|
	\end{center}
This query calculates the expected number of interventions during a trial where the judgment was unanimously favouring the respondents. Other queries can be written in a similar fashion. The detailed results are shown in Table~\ref{tquery10}.
	\begin{table}
	\scriptsize
	\caption{Intervention count comparisons for two extreme scenarios}\label{tquery10}
	\centering
	\begin{tabular}{|l|l|l|l|l|l|}
		\hline
		Expected  &  All &\multicolumn{2}{|c|}{ Unanimous Bench }&\multicolumn{2}{|c|}{Divided Bench}\\\cline{3-6}
		Interventions & Cases &  PP Won & RR Won & PP Won & RR Won \\
		\hline
		Total & 115 & 113 &98 &  125 & 133\\
		\hline
		Petitioner & 55 & 51 & 52& 68 &67\\
		\hline
		Respondent & 60 & 61 &45&57 &65\\
		\hline
		Amicus Curiae & 8 &8 &7& 16&11\\
		\hline
	\end{tabular}
    \end{table}
    \begin{table}
	\scriptsize
	\caption{Percentage distribution of average interventions.}\label{tquery10a}
	\centering
	\begin{tabular}{|l|l|l|l|l|l|}
		\hline
		\multicolumn{2}{|c|}{All cases} &\multicolumn{2}{|c|}{ Unanimous Bench }&\multicolumn{2}{|c|}{Divided Bench}\\\cline{1-6}
		 Winning side & Losing side & Winning side & Losing side & Winning side & Losing side \\
		\hline
        49\%&51\%&45\%&55\%&52\%&48\%\\
		\hline
	\end{tabular}
    \end{table}
We observe that the expected number of interventions decreases for the cases where there is a unanimous decision. We also see that the share of interventions for the winning side for those cases is less than the average share. Similarly, the share of interventions for the losing side is more than the average share. The share in favour to against becomes $45-55$ approximately as compared to $49-51$ for the average over all the cases (see Table~\ref{tquery10a} (Unanimous Bench)). The decline in the total number of interventions indicates that the duration of the trials has decreased (in terms of the number of utterances). In other words, cases tend to go longer when the bench of justices differ in their opinion.\\
	(b) Cases where the judgment was ($5-4$): A similar study was conducted but we now shift our focus to the cases where the bench is most divided, i.e., all those cases which have been decided by a single vote. The queries can be formulated analogously. The results are also shown in Table~\ref{tquery10} (Divided Bench). We see that such trials are longer than the average in terms of the number of utterances. Moreover, the involvement of the justices is higher than the average as they tend to intervene more often to get additional insights into the case which would enable them to make a judgment. Interestingly, we also see a rise in the expected number of interventions for the amicus curiae, showing that justices tend to discuss these cases with the friend of the court to get more clarity on the matter which will help them in reaching a verdict. The intervention share is $52-48$ (see Table~\ref{tquery10a} (Divided Bench)) which is close to the $50-50$ scenario and implies that both sides are equally intervened for these cases.

\subsubsection*{7. Analyzing the behavior of a justice to his/her favorable/unfavorable side.}
	In this query, we study the behavior of the justices for the cases where their viewpoint of the case did not coincide with the majority of the bench. We also compare these results for each side (PP/RR) with their corresponding expected number of interventions by the justices. The results are shown in Table~\ref{tquery13} where $N$ denotes the expected number of interventions, the tuple $(i,j)$ corresponds to the vote of the justice and the outcome of the case, e.g., $(1,0)$ means that the justice voted for the petitioner and the case was decided in favor of the respondents.
   \begin{table}
		\scriptsize
		\caption{Interventions by justice towards his/her favourable/unfavourable side.}
		\label{tquery13}
		\centering
		\begin{tabular}{| c | c | c | c | c | c | c | c |}
			\hline
			\multirowcell{2}{Justice}&\multicolumn{3}{|c|}{Intervening PP}&\multicolumn{3}{|c|}{Intervening RR}\\
			\cline{2-7}
			&N&(0,1)&(1,0)&N&(0,1)&(1,0)\\
			\hline					
			 $J_1$     &   6.2& 10.5& 3.5&  7.9&  7.7& 7.3\\
			 $J_2$      &    0.4 &  0.6 &  0.2 &  0.4 &  0.6 &  0.2\\
			 $J_3$    &    4.2 &  4.2 &  3.9 &   3.3 &  3.8 &  4.5\\
			 $J_4$      &    7.9   &  11.4  &  4.9   &  10.2 &   7.4 &  11.2\\
			 $J_5$       &   6.1   &  11.1 &  2.2  &  6.5 &  4.2 &  12.2\\
			 $J_6$   &    12.6  &  16.4  &  8.1    &  7.7 &   7.7  &  17.6\\
			 $J_7$       &    8.0   &  9.1  &  2.7   &  6.6 & 5.7  &  10.5\\
			 $J_8$     &    5.4   &  6.7  &  3.3    &  8.5 &  4.5  &  16.4\\
			 $J_9$   &    4.5   &  6.4  &  2.9  &  8.5  &  5.7  &  12.0\\
			\hline
		\end{tabular}
	\end{table}
	We can see that for the cases where the justice voted for the petitioner and the case went against them, i.e., $(1,0)$ the expected number of interventions for the petitioner by each justice is smaller as compared to their average interventions, i.e., $N$. In contrast, the expected interventions for the respondents for $(1,0)$ cases is higher than their average interventions for the majority of the justices. A similar trend can be observed in the other direction $(0,1)$ cases. For example, all the justices tend to intervene PP more than their average interventions for these cases. Similarly, the majority of the justices tend to intervene RR less as compared to their averages for cases where justices voted in favor of respondents. Hence, we can conclude that justices tend to intervene more than expected for the advocate who the majority of the bench is convinced with but the justice feels otherwise. Contrarily, the justices intervene less than their averages when the majority of the bench is against them and they favor the advocates.
 
\subsubsection*{8. Analyzing the effect of an intervention by a justice on the advocate.}
    Next, we investigate the impact of interventions on the advocate. Our assumption is that interventions will force the advocate to change the sentiment of his/her speech. Moreover, interventions also indicate that the original argument prepared by the advocate is not going well, and therefore the advocate needs to change how he/she delivers the argument. This would imply more pauses during the argument as the nature of speech shifts from prepared to organic. We tested our hypotheses by formulating the following PRISM queries:
    \begin{center}
    \scriptsize
    	\verb#filter(avg, P=? [ (!"JJ"|"INTERVENING") U<=5 (("PP"|"RR")&"sentiment"#\\\verb#&"pauses")],(("PP"|"RR")&"sentiment"&"pauses"))#
    \end{center}
    It was observed that after the interventions, the chances of an advocate's arguments with negative sentiment and fewer pauses turning into positive sentiment with more pauses are much higher ($0.48$) than the other way round ($0.18$). If we keep the sentiments aside and modify the query to just calculate the effect of interventions on the number of pauses of an advocate, we also observe a similar trend, i.e., the average probability for the transition from fewer pauses to more pauses in $5$ steps is much more ($0.82$) than the other way round ($0.36$). Additionally, we also found out that positive interventions from the justices have a higher chance ($0.13$) of turning an advocate's arguments with negative sentiment and fewer pauses into positive sentiment with more pauses than the other way round ($0.08$). This implies that the impact of a positive intervention is more profound in forcing the advocate to change the sentiment of his/her argument as compared to the negative intervention. If we keep the number of pauses aside and modify the query for just sentiments, a similar trend can be observed, i.e., the average probability for the transition from negative to positive is $0.64$ as compared to $0.23$ for the other way around. These queries inform us about how advocates adapt their arguments during a trial, which was also reported in \cite{relatedcourt1} where authors had claimed the following: "Legal professionals routinely change their language considerably as they move from the monological phases of the trials to the rather informal dialogical phases".

\subsubsection*{9. From an utterance, what is the maximum number of times an advocate is intervened during the next $50$ time units?}
The corresponding PRISM query is as follows:
	\begin{center}
    \scriptsize
		\verb#filter(max,R{"JJ_INTERVENINGside"}=? [C<=50],"PP"|"RR")#\\
	\end{center}
	The result of this query is $28$. More specifically, for the petitioner, the value is $26$ and for the respondents, it is $25$. We convert these queries to experiments over the entire state space of the model to identify the states with maximum values. The results for the top six states are shown below.
	\begin{table}
		\centering
		\scriptsize
		\begin{tabular}{|l|l|}
            \hline
			(PP,ACNO,0,(0,1,1,1,1,0,0,0,0),Normal,SF,Pos,SoU,MP)&25.9\\\hline
            (PP,ACNO,1,(1,1,1,1,1,0,1,1,1),Normal,MS,Neg,LoU,MP)&24.85\\\hline
            (RR,ACNO,0,(1,0,0,1,0,1,0,0,0),Normal,MS,Neg,SoU,LP)&24.76\\\hline
            (RR,ACNO,0,(0,1,0,0,1,0,0,0,0),Normal,SF,Neg,LoU,MP)&24.56\\\hline
            (RR,ACNO,0,(0,0,1,1,0,0,1,0,0),Normal,SF,Neg,LoU,MP)&24.47\\\hline
            (PP,ACNO,1,(1,1,0,1,1,0,1,1,1),Normal,SF,Pos,LoU,LP)&24.17\\\hline
		\end{tabular}
	\end{table}
 
	We observe that except one all the other advocates went on to win the trial. This can be observed from the first index (RR/PP) and the third index (0/1) of the state. We also see that the majority of the utterances have a negative sentiment ($4$ out of $6$) which indicates that justices tend to intervene   more for arguments with a negative sentiment. We also see that none of them is amicus curiae who are considered to be the friends of the court and consulted for their expertise on the matter.
	Additionally, we also computed the following: Within the $60$ time-steps, what is the maximum number of times a particular justice intervenes the advocates from each side. We also studied the voting pattern of the justice for that case and the outcome of the case as well. The PRISM query can be given as follows:
	\begin{center}
    \scriptsize
		\verb|filter(max, R{"jname_INTERVENINGside"}=? [ C<=60 ], x=j)|
	\end{center}
	The maximum values for each justice for both sides, i.e., the petitioner and the respondents are shown below.
			\begin{table}
		\centering
			\scriptsize
		\begin{tabular}{|l|l|l|}
  \hline
			\multirowcell{2}{$J_1$}&(0,(\textbf{0},0,0,1,0,0,0,0,0),InterveningPP,Pos,LoU,MP)&11.6\\ &(0,(\textbf{0},0,1,1,0,0,1,0,0),InterveningRR,Pos,LoU,MP)&9.7\\\hline
			\multirowcell{2}{$J_2$}&(1,(1,\textbf{0},1,1,1,1,1,0,1),InterveningPPRE,Pos,SoU,MP)&4.1\\ &(0,(0,\textbf{0},1,1,0,0,1,0,0),InterveningRR,Pos,SoU,LP)&4.7\\\hline
			\multirowcell{2}{$J_3$}&(0,(0,0,\textbf{0},0,0,0,0,0,IU),InterveningPPRE,Neg,SoU,LP)&7.5\\ &(0,(0,0,\textbf{1},0,0,0,0,1,0),InterveningRR,Neu,SoU,LP)&11.6\\\hline
			\multirowcell{2}{$J_4$}&(1,(1,1,1,\textbf{0},1,1,1,1,IU),InterveningACPP,Neg,LoU,MP)&15.7\\ &(0,(0,0,0,\textbf{1},0,0,0,0,IU),InterveningRR,Pos,LoU,MP)&14.6\\\hline
			\multirowcell{2}{$J_5$}&(0,(0,0,1,1,\textbf{0},1,0,0,0),InterveningPPRE,Neu,SoU,MP)&15.4\\ &(0,(1,0,0,0,\textbf{1},0,0,1,1),InterveningRR,Neg,SoU,LP)&13.2\\\hline
			\multirowcell{2}{$J_6$}&(1,(1,1,0,0,1,\textbf{0},1,1,1),InterveningPPRE,Neu,SoU,MP)&16.5\\ &(0,(0,0,0,1,0,\textbf{1},1,1,0),InterveningRR,Pos,SoU,MP)&13.6\\\hline
			\multirowcell{2}{$J_7$}&(0,(1,1,0,0,0,0,\textbf{0},1,IU),InterveningPP,Neu,LoU,MP)&9.8\\ &(0,(0,0,1,1,0,1,\textbf{1},0,0),InterveningACRR,Neu,LoU,MP)&10.4\\\hline
			\multirowcell{2}{$J_8$}&(0,(1,0,0,1,1,0,0,\textbf{0},0),InterveningPP,Neu,SoU,MP)&11.2\\ &(0,(1,1,0,0,1,0,0,\textbf{1},0),InterveningRR,Pos,SoU,MP)&12.8\\\hline
			\multirowcell{2}{$J_9$}&(0,(1,0,0,1,1,0,0,0,\textbf{0}),InterveningPP,Pos,LoU,MP)&9.2\\
            &(0,(0,1,1,1,1,0,0,0,\textbf{0}),InterveningRR,Neu,SoU,MP)&11.9\\\hline
		\end{tabular}
	\end{table}
    
    For the respondents, we see that in all the cases they went on to win the case (see the second row for each justice). Interestingly, except $J_1, J_2$ and $J_9$, all the other justices have voted against the advocates for cases where they intervened the most in $60$ steps.
    This shows that an attempt was made by the justices to convey his/her concerns about the argument of the advocate which otherwise has convinced the majority of the justices of the bench.
    
    Interestingly, we see that all the justices went on to vote against the petitioner in the states where they intervened the most (see the first row and the highlighted $0$ for each justice). Collectively, in the majority of the cases, the petitioner has lost the case ($15$ out of $18$ rows). By observing the values of both PP and RR for each justice (rightmost column), it can be concluded that no justice goes out of his/her way while intervening the advocates of either side.

\subsubsection*{10. Analyzing the rebuttal arguments.}
	Rebuttal arguments are usually not intervened much by the justices. In fact, in many cases ($42\%$) they go completely unintervened by the justices which is evident from the following PRISM query:
 \begin{center}
 \scriptsize
 \verb|P=?[(!"InterveningPPRE")U("FINAL")]|=$0.42$
 \end{center}
 Having said that, we see frequent interventions occurring in some of the cases. We investigate this further by identifying and analyzing the corresponding states.
	We run the following PRISM query: \verb|filter(max,R{"r_INTERVENING"}|\\\verb|r=?[F "FINAL"],"REBUTTAL"&(x=j))|. The results are as follows.
	\begin{table}
		\centering
		\scriptsize
		\begin{tabular}{|l|l|}\hline
	(PP,ACNO,1,(1,1,0,0,1,0,1,1,1),Rebuttal,MS,Pos,LoU,MP)&24.0\\\hline
	(PP,ACNO,0,(0,0,1,1,0,1,0,0,0),Rebuttal,SF,Pos,SoU,LP)&17.33\\\hline
	(PP,ACNO,1,(1,1,1,1,1,0,0,1,1),Rebuttal,MS,Neu,LoU,MP)&16.42\\\hline
    (PP,ACNO,1,(1,1,0,1,1,0,1,1,1),Rebuttal,SF,Neu,SoU,LP)&15.25\\\hline
    (PP,ACNO,1,(1,0,1,1,0,1,1,0,1),Rebuttal,SF,Pos,SoU,LP)&15.0\\\hline
    \end{tabular}
    \end{table}
	
    We observe that in $4$ out of $5$ instances, the petitioner wins the case. Since the rebuttal is the last concluding argument during a proceeding, it indicates that the bench has already made up its mind about the case and the justices would like to clarify some reservations or provide comments about the case before closing the proceeding. If we conduct a similar experiment where the role is justice, we can identify the states for each justice which resulted in the most interventions during the rebuttals. The corresponding PRISM query is as follows: \verb|filter(max,R{"r_INTERVENING"}=?[F "FINAL"],"INTERVENINGPPRE"&(x|\verb|=j))|. The results are shown below.
		\begin{table}
		\centering
		\scriptsize
		\begin{tabular}{|l|l|}\hline
			($J_6$,1,(1,1,0,0,1,\textbf{0},1,1,1),SF,Neu,SoU,MP)&25.0\\\hline
            ($J_5$,0,(0,0,1,1,\textbf{0},1,0,0,0),MS,Neu,SoU,MP)&17.33\\\hline
            ($J_3$,1,(1,1,\textbf{0},1,1,0,1,1,1),MS,Neu,SoU,MP)&17.25\\\hline
            ($J_6$,1,(1,1,1,1,1,\textbf{0},0,1,1),MS,Neu,SoU,MP)&16.75\\\hline
            ($J_5$,0,(0,0,1,1,\textbf{0},1,0,0,0),MS,Neg,SD,MP)&16.67\\\hline
            ($J_6$,1,(1,0,1,1,0,\textbf{1},1,0,1),MS,Neu,SD,MP)&15.5\\\hline
		\end{tabular}
	\end{table}
	
    We see that $J_6$ and $J_5$ are involved in majority of the instances which resulted in the most interventions during the rebuttals. We also see that for the top $5$ instances, the corresponding justice's vote went against the petitioner. Interestingly, the sentiment is neutral for all the cases.
 
\subsubsection*{11. Analyzing those events that led to the multiple justices jointly intervening in an argument.}
	In the dataset and in our previous queries, it is rare to find instances where multiple justices jointly intervene in an argument. In this query, we try to study scenarios where justices have jointly intervened in an argument. We use the following query to calculate the expected length of an inter-jury discussion following an argument by an advocate.
	\begin{center}
		\scriptsize
		\verb#filter(state,R{"n_step"}=?[(X("JJ"))U(X(!("JJ")))],(("PP"|"RR")&(x=j)))#
	\end{center}
	We sort the experiment results for the states in descending order, and the results are shown below.
	\begin{table}
	\centering
	\scriptsize
	\begin{tabular}{|l|l|}\hline
	(RR,ACNO,0,(1,0,0,1,0,1,0,0,0),MS,Neg,SoU,LP)&5.0\\\hline
    (RR,ACNO,0,(0,1,0,0,0,0,0,1,IU),SF,Pos,LoU,LP)&5.0\\\hline
    (RR,ACNO,1,(1,1,0,0,1,1,0,1,1),SF,Neu,SoU,LP)&5.0\\\hline
    (PP,ACNO,1,(1,1,1,1,1,0,1,1,1),SF,Pos,LoU,MP)&5.0\\\hline
    (PP,ACNO,0,(0,1,0,0,0,0,0,0,0),SF,Pos,SoU,MP)&4.75\\ \hline
    (PP,ACNO,0,(0,0,0,0,0,0,0,0,0),LG,Neu,SoU,MP)&4.5\\\hline
	\end{tabular}
	\end{table}
	
    We observe that in all the instances at least $6$ out of $9$ justices have voted for the same side. This indicates that scenarios, where at least $6$ like-minded justices are present, are the ones where the joint intervention takes place. Next, we try to investigate this further by asking the following question: in a joint intervention scenario, are all the justices who are involved in the intervention like-minded or not? The query can be modified as follows:
	\begin{center}
		\scriptsize
		\verb|filter(max, R{"n_step"}=? [ ("JJ"&"INTERVENING") U !("JJ") ],(("JJ"&"INTERVENING")&(x=j)))|
	\end{center}
	The states with the maximum values and the list of all the justices involved in the intervention are shown in the table below.
	\begin{table}
	\centering
	\scriptsize
	\begin{tabular}{|l|l|l|}\hline
	($J_7$,0,(\textbf{0},1,0,0,0,0,\textbf{0},\textbf{0},0),InterveningPP,Neu,SoU)& 12.0 &$J_8$, $J_1$\\\hline
	($J_7$,1,(\textbf{1},0,\textbf{1},1,0,1,\textbf{1},0,1),InterveningRR,Pos,SoU)& 8.0 &$J_1$, $J_3$\\\hline
	($J_8$,1,(\textbf{1},1,1,1,1,1,1,\textbf{1},\textbf{1}),InterveningRR,Neu,SoU)&6.37&$J_9$, $J_1$\\\hline
    ($J_8$,0,(0,0,0,0,0,0,\textbf{0},\textbf{0},\textbf{0}),InterveningPP,Neg,SoU)&6.0&$J_7$,$J_9$\\\hline
    ($J_9$,1,(1,0,1,1,1,1,\textbf{1},0,\textbf{1}),InterveningPP,Neu,SoU)&6.0&$J_7$\\\hline
    \end{tabular}
	\end{table}

    It can be observed that in all the instances/rows, justices who are involved in joint intervention are like-minded, e.g., for the first instance/row, $J_7, J_8$ and $J_1$ are involved and all these justices are like-minded (see the three highlighted 0's in the 9-tuple). Similarly, for the third instance/row, $J_8, J_9$, and $J_1$ are involved and these justices are like-minded. Additionally, we also see the trend that like-minded justices who jointly intervene, tend to intervene the advocate of the opposite side. We also see that the case outcome is also aligned with their inclinations. Finally, since these are joint interventions, they are all small utterances, i.e., SoU.
 
\subsubsection*{12. Analyzing the sequences where a justice and an advocate had an extended dialogue.}
	In this query, we tried to look at the states that led to an extended dialogue between a justice and an advocate. Identifying such states would provide interesting insights about the nature of the arguments by an advocate, his/her side, and the sentiment that leads to such interactions. Additionally, the justices involved in such interaction can also be identified. We use the following query to obtain the maximum expected length of dialogues between a justice and an advocate:
	\begin{center}
		\scriptsize
		\verb#filter(max, R{"n_step"}=? [ ("RR"|"PP"|"jname") U ("JJ"&!("jname")) ], ("RR"|"PP"))#
	\end{center} 
    The average maximum expected length of dialogues is $12$. For justice $J_1$, the result is not defined in the query. This is due to the following reason: this is a reachability property on rewards, and if the probability of satisfying the query is less than $1$, then the expected reward is infinite. Next, we modify this query and run an experiment to find out which states led to such scenarios. The states which led to the longest one-on-one conversations between the justice and an advocate is listed below with the names of the corresponding justices and the lengths of the dialogue.
	\begin{table}
		\centering
    \scriptsize
	\begin{tabular}{|l|l|l|}\hline
		(PP,ACNO,1,(1,0,1,1,0,\textbf{1},1,0,1),Rebuttal,SF,Pos,SoU,LP)&29.0&$J_6$\\\hline
		(PP,ACNO,0,(0,0,1,1,\textbf{0},1,0,0,0),Rebuttal,MS,Pos,SoU,MP]&20.33&$J_5$\\\hline
		(PP,ACNO,1,(1,1,1,\textbf{1},1,1,1,1,IU),Rebuttal,MS,Neu,SoU,MP)&19.0&$J_4$\\\hline
		(PP,ACNO,1,(1,1,1,1,1,\textbf{0},0,1,1),Rebuttal,MS,Neu,LoU,MP)&17.14&$J_6$\\\hline
		(PP,ACNO,1,(1,0,1,1,\textbf{1},1,0,0,1),Rebuttal,SF,Neg,LoU,MP)&15.0&$J_5$\\\hline
	\end{tabular}
	\end{table}	
	
    Another observation is that all these conversations took place with the petitioner side, and the petitioner won in all those cases except one, i.e., the second row. Additionally, we see that the five most lengthy dialogues occurred during the rebuttals, implying that the intervention must have started in the earlier part of the rebuttal argument. Interestingly, as mentioned earlier (query $10$), for $42\%$ of the cases, rebuttals go unintervened, but the result of this query shows that for certain cases rebuttals may result in long one-on-one conversations between a justice and an advocate. A possible explanation is that although the justices have already made up their minds about these cases, they would still like to clarify some points before coming to the final conclusion.

\section{Conclusion \& Future Work}\label{sec:conc}
This paper demonstrated that probabilistic model checking can be very effective for modeling, analysis, and verification of temporal dynamics of human communication in the judiciary. We built a DTMRM using the transcripts of the oral arguments of cases from the United States Supreme Court. Interesting queries were specified using PCTL and PCTL with rewards and verified using the PRISM model checker. Our results showed that probabilistic model checking can uncover hidden patterns, identify trends and provide valuable feedback to the justices, advocates, and other stakeholders of the judiciary. In the future, we plan to investigate the following: 1) generalize this approach so that it can be applied to all the cases from the U.S. Supreme Court. This would require applying aggressive minimization techniques to tackle the state space explosion problem. 2) Add more interesting linguistic markers, e.g., `Dialogue-Acts', and `Discourse' etc. for constructing the Markovian model. 3) We also plan to apply this approach to other group interaction datasets, e.g., \cite{nonverbal,dataset1}.    

\bibliographystyle{splncs04}
\bibliography{supremebib}

\begin{thebibliography}{10}
\providecommand{\url}[1]{\texttt{#1}}
\providecommand{\urlprefix}{URL }
\providecommand{\doi}[1]{https://doi.org/#1}

\bibitem{website-ckit}
\url{https://convokit.cornell.edu/documentation/supreme.html}

\bibitem{Andova}
Andova, S., Hermanns, H., Katoen, J.: Discrete-time rewards model-checked. In:
  {FORMATS}. pp. 88--104. LNCS 2791, Springer (2003)

\bibitem{oana1}
Andrei, O., Murray, G.: Interpreting models of social group interactions in
  meetings with probabilistic model checking. In: Proceedings of the Group
  Interaction Frontiers in Technology Workshop, GIFT@ICMI 2018, Boulder, CO,
  USA, October 16, 2018. pp. 5:1--5:7 (2018)

\bibitem{relatedcourt1}
Aronsson, K., J{\"o}nsson, L., Linell, P.: The courtroom hearing as a middle
  ground: Speech accommodation by lawyers and defendants. Journal of Language
  and Social Psychology  \textbf{6},  115--99 (1987)

\bibitem{KB08}
Baier, C., Katoen, J.P.: Principles of Model Checking. MIT Press (2008)

\bibitem{relatedcourt3}
Benus, S., Levitan, R., Hirschberg, J.: Entrainment in spontaneous speech: The
  case of filled pauses in supreme court hearings. In: 2012 IEEE 3rd
  International Conference on Cognitive Infocommunications (CogInfoCom). pp.
  793--797 (2012)

\bibitem{nltk}
Bird, S., Klein, E., Loper, E.: Natural Language Processing with Python.
  O'Reilly (2009)

\bibitem{dataset1}
Chang, J.P., Chiam, C., Fu, L., Wang, A.Z., Zhang, J.,
  Danescu{-}Niculescu{-}Mizil, C.: Convokit: {A} toolkit for the analysis of
  conversations. In: Proceedings of the 21th Annual Meeting of the Special
  Interest Group on Discourse and Dialogue, SIGdial 2020, 1st virtual meeting,
  July 1-3, 2020. pp. 57--60. Association for Computational Linguistics (2020)

\bibitem{silent_tom2}
Cole, D., Biskupic, J., de~Vogue, A.: Clarence thomas breaks silence on bench
  during supreme court’s first remote oral argument  (May 2020),
  \url{https://edition.cnn.com/2020/05/04/politics/clarence-thomas-question/index.html}

\bibitem{cornwell}
Cornwell, B.: Social Sequence Analysis: Methods and Applications. Structural
  Analysis in the Social Sciences, Cambridge University Press (2015)

\bibitem{dataset}
Danescu-Niculescu-Mizil, C., Lee, L., Pang, B., Kleinberg, J.: Echoes of power:
  {Language} effects and power differences in social interaction. In:
  Proceedings of WWW. pp. 699--708 (2012)

\bibitem{STORM}
Dehnert, C., Junges, S., Katoen, J., Volk, M.: A storm is coming: {A} modern
  probabilistic model checker. In: {CAV}. pp. 592--600. LNCS 10427, Springer
  (2017)

\bibitem{relatedcourt5}
Dietrich, B.J., Enos, R.D., Sen, M.: Emotional arousal predicts voting on the
  u.s. supreme court. Political Analysis  \textbf{27}(2),  237–243 (2019)

\bibitem{PCTL2}
Hansson, H., Jonsson, B.: A logic for reasoning about time and reliability.
  Formal Asp. Comput.  \textbf{6}(5),  512--535 (1994)

\bibitem{relatedcourt4}
Hawes, T., Lin, J., Resnik, P.: Elements of a computational model for
  multi-party discourse: The turn-taking behavior of supreme court justices.
  Journal of the American Society for Information Science and Technology
  (2009)

\bibitem{storm1}
Hensel, C., Junges, S., Katoen, J., Quatmann, T., Volk, M.: The probabilistic
  model checker storm. Int. J. Softw. Tools Technol. Transf.  \textbf{24}(4),
  589--610 (2022)

\bibitem{MRMC}
Katoen, J., Khattri, M., Zapreev, I.S.: A markov reward model checker. In:
  Second International Conference on the Quantitative Evaluaiton of Systems
  {(QEST} 2005), 19-22 September 2005, Torino, Italy. pp. 243--244. {IEEE}
  Computer Society (2005)

\bibitem{dtmc1}
Kemeny, J.G., Snell, J.L., et~al.: Finite {M}arkov chains, vol.~356. van
  Nostrand Princeton, NJ (1960)

\bibitem{rewards2}
Kwiatkowska, M.Z., Norman, G., Parker, D.: Stochastic model checking. In:
  Formal Methods for Performance Evaluation, 7th International School on Formal
  Methods for the Design of Computer, Communication, and Software Systems,
  {SFM} 2007, Bertinoro, Italy, May 28-June 2, 2007, Advanced Lectures. pp.
  220--270 (2007)

\bibitem{PRISM1}
Kwiatkowska, M.Z., Norman, G., Parker, D.: {PRISM} 4.0: Verification of
  probabilistic real-time systems. In: {CAV}. pp. 585--591. LNCS 6806, Springer
  (2011)

\bibitem{silent_tom3}
Liptak, A.: Justice clarence thomas, long silent, has turned talkative  (May
  2021),
  \url{https://www.nytimes.com/2021/05/03/us/politics/clarence-thomas-supreme-court.html}

\bibitem{murrayold}
Murray, G.: Markov reward models for analyzing group interaction. In:
  Proceedings of the 19th {ACM} International Conference on Multimodal
  Interaction, {ICMI} 2017, Glasgow, United Kingdom, November 13 - 17, 2017.
  pp. 336--340 (2017)

\bibitem{nonverbal}
Sanchez-Cortes, D., Aran, O., Mast, M.S., Gatica-Perez, D.: A nonverbal
  behavior approach to identify emergent leaders in small groups. IEEE
  Transactions on Multimedia  \textbf{14}(3),  816--832 (2012)

\bibitem{Sharma1}
Sharma, A.: Weighted probabilistic equivalence preserves {\(\omega\)}-regular
  properties. In: {MMB} {\&} {DFT}. pp. 121--135. LNCS 7201, Springer (2012)

\end{thebibliography}
\end{document}